\documentclass[journal]{IEEEtran}
\usepackage{cite}
  \usepackage[pdftex]{graphicx}
\usepackage{amsmath,bm}
\usepackage[ruled,linesnumbered]{algorithm2e}

\usepackage{array}
\usepackage{subfigure}

\usepackage{color}

\begin{document}

\title{Cooperative NOMA-Based User Pairing for URLLC: A Max-Min Fairness Approach}

\author{Fateme~Salehi,~
        Naaser~Neda,~
        Mohammad-Hassan~Majidi,~and~Hamed~Ahmadi% <-this % stops a space
\thanks{F. Salehi is with the Faculty of Electrical and Computer Engineering, University of Birjand, Birjand, Iran (e-mail: f.salehi@birjand.ac.ir).}% <-this % stops a space
\thanks{N. Neda {\emph{(Corresponding author)}} is with the Faculty of Electrical and Computer Engineering, University of Birjand, Birjand, Iran (e-mail: nneda@birjand.ac.ir).}% <-this % stops a space
\thanks{M.-H. Majidi is with the Faculty of Electrical and Computer Engineering, University of Birjand, Birjand, Iran (e-mail: m.majidi@birjand.ac.ir).}% <-this % stops a space
\thanks{H. Ahmadi is with the Department of Electronic Engineering,  University of York, United Kingdom (e-mail: hamed.ahmadi@york.ac.uk).}% <-this % stops a space
\thanks{Part of this paper has been presented in EuCNC 2021 \cite{eucnc_conf}.}
\thanks{Manuscript received January 19, 2021; revised May 3, 2021 and September 1, 2021; accepted September 25, 2021.}
}

\maketitle

\begin{abstract}
In this paper, cooperative non-orthogonal multiple access (C-NOMA) is considered in short packet communications with finite blocklength (FBL) codes. The performance of a decode-and-forward (DF) relaying along with selection combining (SC) and maximum ratio combining (MRC) strategies at the receiver side is examined. We explore joint user pairing and resource allocation to maximize fair throughput in a downlink (DL) scenario. In each pair, the user with a stronger channel (strong user) acts as a relay for the other one (weak user), and optimal power and blocklength are allocated to achieve max-min throughput. To this end, first, only one pair is considered, and optimal resource allocation is explored. Also, a suboptimal algorithm is suggested, which converges to a near-optimal solution. Finally, the problem is extended to a general scenario, and a suboptimal C-NOMA-based user pairing is proposed. Numerical results show that the proposed C-NOMA scheme in both SC and MRC strategies significantly improves the users’ fair throughput compared to the NOMA and OMA. It is also investigated that the proposed pairing scheme based on C-NOMA outperforms the Hybrid NOMA/OMA scheme from the average throughput perspective, while the fairness index degrades slightly.
\end{abstract}


%\IEEEpeerreviewmaketitle

\section{Introduction}\label{sec.intro}

\IEEEPARstart{T}{he} ever-increasing new demands such as tactile internet, high-resolution video streaming, virtual/augmented reality, autonomous vehicles, etc., with various requirements, may be somewhat challenging in terms of reliability and latency. Unlike most of the existed mobile networks designed for traditional mobile broadband (MBB) services, Internet-of-Things (IoT) attempts to connect plentiful devices with the least human intervention. IoT applications are divided into massive machine-type communications (mMTC) and ultra-reliable low-latency communications (URLLC). The first one consists of many low-cost devices with massive connections and high battery lifetime requirements. On the other hand, URLLC requirements are most related to mission-critical services in which the importance of uninterrupted and robust data exchange is far greater than anything else.

Short packets with FBL codes are considered to reduce the transmission delay and support low-latency communication. In the FBL regime communication, in contrast to Shannon's capacity for infinite blocklength, decoding error probability at the receiver is not negligible owing to short blocklength \cite{IT_1}. Polyanskiy et al. succeeded in deriving an exact approximation of the FBL regime’s information rate at the AWGN channel \cite{IT_2}. Following that, research in this context developed to MIMO channel with quasi-static fading \cite{IT_3} and a quasi-static fading channel with retransmissions \cite{IT_4}. Furthermore, the effect of short packets on the spectrum sharing, and scheduling of delay-sensitive packets was considered in \cite{IT_5} and \cite{IT_6}, respectively. In \cite{IT_7}, massive MIMO adoption to maximize the achievable uplink data rate for industrial applications was advocated for both MRC and zero-forcing (ZF) receivers. In \cite{IT_8}, the resource allocation for a secure mission-critical IoT communication system was studied under finite blocklength, and two optimization problems with the aim of weighted throughput maximization and total transmit power minimization were addressed. The authors in \cite{IT_9} proposed a cross-layer framework for optimizing user association, packet offloading rates, and bandwidth allocation for mission-critical IoT scenarios.

The NOMA performance in the FBL regime was studied in \cite{NOMA_1,NOMA_2,NOMA_3,NOMA_4}. In \cite{NOMA_1}, optimal power and blocklength allocation was considered in a high signal-to-noise ratio (SNR) scenario, and the amount of NOMA transmission delay reduction was determined compared to OMA in a closed-form. In \cite{NOMA_2}, transmission rate and power allocation of the NOMA scheme were optimized to maximize the effective throughput of the strong user, while the throughput of the other user was guaranteed at a certain level. The transmitter’s energy with a hybrid transmission scheme that combines the time division multiple access (TDMA) and NOMA was minimized in \cite{NOMA_3} subject to heterogeneous latency constraints at receivers. In \cite{NOMA_4}, an optimal power allocation algorithm was proposed to achieve max-min throughput under energy, reliability, and delay constraints in a DL-NOMA transmission and compared with its optimal OMA counterpart.

Relaying is a well-known technique to increase capacity and reliability. In \cite{relay_1}, relaying performance in the FBL regime was studied, and its advantages over the direct transmission were investigated. The throughput and effective capacity of a relaying system in the FBL regime were obtained in \cite{relay_2} at the presence of a quasi-static fading channel and some assumptions on average channel state information (CSI) at the transmitter. In \cite{relay_3}, under the assumption of outdated CSI at the source, the authors maximized the FBL throughput of a two-hop relaying system while guaranteeing a reliability constraint. 

Ding et al. in \cite{C-NOMA_1} proposed the cooperative NOMA transmission scheme, a cooperative relaying technique in the NOMA system which fully exploits the prior knowledge available by applying the successive interference cancellation (SIC) strategy. Followed by that, they introduced a two-stage relay selection strategy in the C-NOMA network \cite{C-NOMA_2}. In \cite{C-NOMA_3}, a buffer-aided C-NOMA scheme, where the intended users are equipped with buffers for cooperation, was proposed to adaptively select a direct or cooperative transmission mode, based on the instantaneous CSI and the buffer state. In \cite{C-NOMA_4}, the authors proposed threshold-based selective C-NOMA, where the strong user forwards the symbols of weak user only if the signal-to-interference-plus-noise ratio (SINR) is greater than the pre-determined threshold value, to increase the data reliability of conventional C-NOMA networks. In \cite{C-NOMA_5}, the authors investigated C-NOMA scheme in short-packet communications with flat Rayleigh fading channels and derived the average block error rate (BLER) of the central user and the cell-edge user theoretically for both SC and MRC strategies.

Optimization problems of average throughput and max-min throughput were studied in \cite{relay_7} with power and blocklength allocation between users under delay and consumed energy constraints by full search method with high complexity, but users’ reliability was not guaranteed. In \cite{relay_8}, Ren et al. considered optimal power and blocklength allocation in OMA, NOMA, relaying, and C-NOMA transmissions schemes to minimize the weak user’s decoding error probability; meanwhile, the reliability of the strong user’s performance was guaranteed at a certain level. Both \cite{relay_7} and \cite{relay_8} have considered a two-user scenario.

In \cite{cluster_2}, a joint user pairing and power allocation problem was explored in a DL-NOMA network to optimize the achievable sum rate with minimum rate constraint for each user. In \cite{cluster_3}, a two-step user-pairing scheme maximizing the achievable diversity gain for an OFDM-based relaying NOMA system with fixed-rate transmission was proposed by selecting one near user and one far user for each subcarrier where far users cannot communicate with the base station (BS) directly. Zhang et al. in \cite{cluster_4} investigated a distance-based user pairing in the C-NOMA network, where the locations of the source and typical user are fixed, and the candidate users for pairing follow the distribution of homogeneous Poisson Point Process, and two close-to-user pairing  and close-to-source pairing schemes were proposed. The authors in \cite{cluster_5} considered user pairing policy and power control scheme jointly in a DL C-NOMA system, where the objective is maximizing the achievable sum-rate of the whole system while guaranteeing a certain quality of service (QoS) for all users. In \cite{cluster_6}, a joint user pairing and subchannel assignment algorithm was proposed in a DL C-NOMA network that pairs a strong user with a weak user and assigns them a subchannel simultaneously, while a Stackelberg game is employed to allocate power among the users by the BS. All of the works in \cite{cluster_2,cluster_3,cluster_4,cluster_5,cluster_6} investigate the user pairing problem in conventional communication with infinite blocklength. Moreover, the last two works do not take into consideration the geometric distance between the paired nodes in the C-NOMA scheme.

In this work, we consider a DL C-NOMA network in the short packet communications scenario. It is assumed that the paired users, their channel gain difference is high. The strong user, which performs SIC and detects the weak user’s data, acts as a relay. The weak user, which receives its data via BS and relay separately can implement SC or MRC to detect its data. To the best of the authors’ knowledge, this is the first work to address the problem of joint user pairing, blocklength and power allocation in a critical IoT scenario.

Our main contributions in this work are summarized as follows:
\begin{enumerate}
  \item We obtain each user’s decoding error probability in the C-NOMA transmission scheme for both SC and MRC protocols with the CSI at the transmitter (CSIT) assumption. The MRC protocol is considered for the first time in the FBL regime with different blocklengths.
  \item To guarantee the quality of service (QoS) of the weak user and to improve fairness, joint power and blocklength optimization is done in both NOMA and relay phases to maximize the minimum throughput of two users in different combining scenarios, under latency, reliability, and energy constraints.
  \item A suboptimal solution with near-optimal performance is proposed to decrease the complexity of the optimal resource allocation, and their computational complexity is determined.
  \item	The problem is extended to a multi-user scenario, and a novel joint suboptimal C-NOMA-based user pairing and resource allocation scheme is proposed. Meanwhile, the simulation results show its comparable performance to the exhaustive-search optimal algorithm.
\end{enumerate}
The remainder of this paper is organized as follows. In Section \ref{sec.prelimin}, the system model and direct transmission analysis in the FBL regime are presented. Performance analysis of the C-NOMA transmission consist of SC and MRC strategies is provided in Section \ref{sec.perform}. Problem formulation with a focus on one pair is considered in Section \ref{sec.formulation}. The optimal and one suboptimal solution are proposed for the problem in Section \ref{sec.solving}. The problem is extended to a multi-user scenario and, one user pairing scheme is proposed in Section \ref{sec.extension}. Numerical results are presented in Section \ref{sec.results}. Finally, Section \ref{sec.conclusion} concludes the paper.

\section{Preliminaries Issues}\label{sec.prelimin}
\subsection{System Model}
\begin{figure}
\centering
    \begin{subfigure}
        \centering
        \includegraphics[width=1\columnwidth,trim={6cm 4.5cm 5cm 14.7cm},clip]{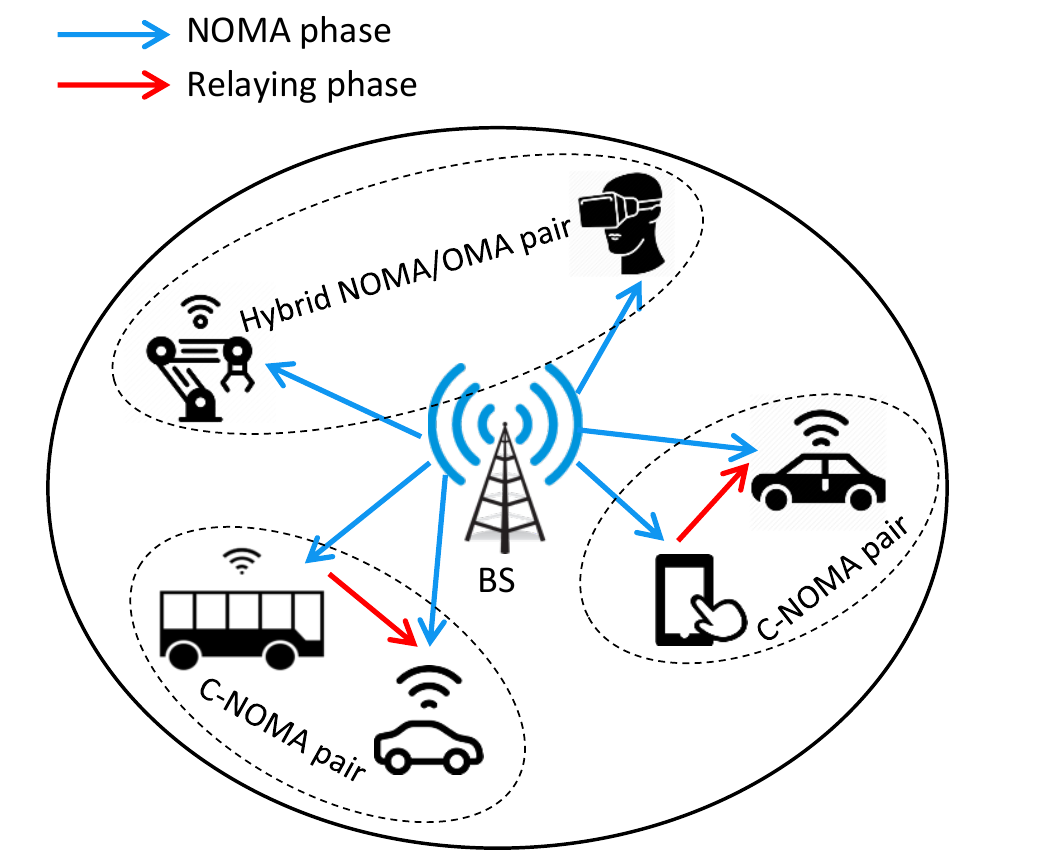}\\ \footnotesize{(a)}
        %\caption{~}
        %\label{fig.sys}
    \end{subfigure}
    \vfill
    \begin{subfigure}
        \centering
        \includegraphics[width=1\columnwidth,trim={6cm 14.7cm 6cm 10cm},clip]{figs.pdf}\\ \footnotesize{(b)}
        %\caption{~}
        %\label{fig.frm}
    \end{subfigure}
    \vfill
    \caption{(a) system model, (b) frame structure.}
    \label{fig.system}
\end{figure}

As shown in Fig. \ref{fig.system}(a), the URLLC users with different QoS requirements are paired into disjoint clusters. For simplicity, we first just focus on one pair. Section \ref{sec.extension} will provide more details of user pairing. Here we consider a cooperative relaying scenario in a DL system with one BS and two NOMA users in each C-NOMA pair.  In phase I, i.e., NOMA phase, BS transmits a NOMA frame of length $m^\textrm{I}$  symbols, which consists of two users’ data ($N_1$  bits, user 1’s data and $N_2$  bits, user 2’s data). User 1, the strong user, performs the SIC technique and decodes user 2’s data and sends that to user 2 in a frame of length $m^\textrm{II}$ symbols in phase II, i.e., relaying phase. The instantaneous channel coefficients of BS-user 1, BS-user 2, and user 1-user 2 links representing small scale fading and large scale fading are denoted as $h_1$, $h_2$, and  $h_{1,2}$, respectively. It is assumed that the channels are quasi-static Rayleigh fading. Hence, they are constant during one frame and vary independently from one frame to the next one. 

According to the power domain NOMA principle, in a two-user scenario, BS transmits $\sum_{i=1}^{2} \sqrt{p_i^\textrm{I}}x_i$, where $x_i$  is the message of user $i$, $i \in \{1,2\}$, and $p_i^\textrm{I}$  refers to the allocated power of user $i$ in phase I. So, the received signal at user $i$ is given by  $y_i^\textrm{I}=(\sqrt{p_1^\textrm{I}}x_1+\sqrt{p_2^\textrm{I}}x_2)h_i+n_i$, where  $n_i$ is the complex additive white Gaussian noise with variance  $\sigma^2$. Without loss of generality, it is assumed that  $|h_1|^2>|h_2|^2$, and more power should be allocated to user 2. Therefore, user 1 can perform the SIC technique to remove the interference, while user 2 suffers from the interference and cannot cancel it. If $x_2$  is decoded correctly by user 1, it is re-encoded and transmitted (denoted by $\sqrt{p_2^\textrm{II}}x_2^\prime$). \footnote{One should notice that $x_2$  is user 2’s data with rate  ${N_2}/{m^\textrm{I}}$, while $x_2^\prime$  is the same data with rate  ${N_2}/{m^\textrm{II}}$.} Consequently, the received signal at user 2 in the relaying phase is  $y_2^\textrm{II}=\sqrt{p_2^\textrm{II}}x_2^\prime~h_{1,2}+n_{1,2}$. Let  $p_2^\textrm{II}$ show the allocated power to user 2 by the relay (user 1) in phase II, and  $n_{1,2}$ is the complex additive white Gaussian noise with variance $\sigma^2$ . To implement this scheme, user 1 must know whether SIC is successful or not. To this end, we suppose that BS sends the channel coding information of both user 1 and user 2 to user 1 via an error-free dedicated channel. The channel coding can help to diagnose whether the decoded data is correct or not. Thus, user 1 knows whether the SIC is successful or not \cite{relay_8}.

\subsection{Direct Transmission Analysis in the FBL Regime}
According to \cite{IT_2}, the achievable data rate $R$  for a finite blocklength of $m$  symbols $(m\geq100 )$, and an acceptable BLER $\varepsilon$ , has an exact approximation as
\begin{equation}
    \label{eq.rate}
    R\approx C-\sqrt{\frac{V}{m}}\frac{Q^{-1}(\varepsilon)}{\ln2}
\end{equation}
where $C=\log_2(1+\gamma)$  is the Shannon capacity, $\gamma$  is the SNR/SINR ratio, $Q^{-1}(\cdot)$  refers to the inverse Gaussian Q-function $Q(x)=\frac{1}{\sqrt{2\pi}}\int_{x}^{\infty} e^{-\frac{t^2}{2}} \,dt$, and $V=1-(1+\gamma)^{-2}$  is the channel dispersion. In the FBL regime, even with perfect CSI, the transmission is not error-free and the decoding error probability is given by
\begin{equation}
    \label{eq.error}
    \varepsilon\approx Q(f(\gamma,R,m)).
\end{equation}
where $f(\gamma,R,m)\overset{\Delta}{=}\frac{(C-R)\ln2}{\sqrt{V/m}}$.

\section{Performance Analysis of C-NOMA Transmission}\label{sec.perform}
It is assumed that the receivers have access to perfect CSI, and BS and each of the users have one antenna. Also, user 2 can employ various combining strategies, including SC and MRC. In phase I, user 2 directly detects $x_2$  by considering $x_1$  as interference. The decoding error probability of $x_2$  at user 2 in phase I is denoted by $\varepsilon_{2,2}^\textrm{I}$ , which is approximated based on \eqref{eq.error} by
\begin{equation}
    \label{eq.error22I}
     \varepsilon_{2,2}^\textrm{I}\approx Q(f(\gamma_{2,2}^\textrm{I},R_{2,2}^\textrm{I},m^\textrm{I}))
\end{equation}
where $\gamma_{2,2}^\textrm{I}=p_2^\textrm{I}|h_2|^2/(p_1^\textrm{I}|h_2|^2+\phi\sigma^2)$  and $R_{2,2}^\textrm{I}=N_2/m^\textrm{I}$ are the received SINR and the achievable rate of user 2 related to detecting $x_2$  in phase I, respectively. $\phi>1$ reflects the SNR/SINR loss due to the imperfect CSI. \footnote{Invoking \cite{IT_10}, the effect of channel estimation error on data rate can be equivalent to noise enhancement, which depends on the velocity of the devices. For devices with slow or medium velocity, $\phi$ is close to 1.} Since  $x_2$ is detected directly, $\varepsilon_{2,2}^\textrm{I}$  is the overall error probability of user 2 in phase I, i.e., $\varepsilon_2^\textrm{I}=\varepsilon_{2,2}^\textrm{I}$ . On the opposite, user 1 performs SIC, meaning it first decodes $x_2$  while treats $x_1$  as interference. Similarly, the decoding error probability of $x_2$  at user 1 in phase I, which is denoted by $\varepsilon_{1,2}^\textrm{I}$, is approximated as
\begin{equation}
    \label{eq.error12I}
     \varepsilon_{1,2}^\textrm{I}\approx Q(f(\gamma_{1,2}^\textrm{I},R_{1,2}^\textrm{I},m^\textrm{I}))
\end{equation}
where $\gamma_{1,2}^\textrm{I}=p_2^\textrm{I}|h_1|^2/(p_1^\textrm{I}|h_1|^2+\phi\sigma^2)$  and $R_{1,2}^\textrm{I}=N_2/m^\textrm{I}$  are the received SINR and the achievable rate of user 1 related to detecting $x_2$  in phase I, respectively. If user 1 decodes and removes $x_2$  successfully, then $x_1$  can be detected without interference. Accordingly, the decoding error probability of $x_1$  at user 1 in phase I, i.e., $\varepsilon_{1,1}^\textrm{I}$ , is denoted by
\begin{equation}
    \label{eq.error11I}
     \varepsilon_{1,1}^\textrm{I}\approx Q(f(\gamma_{1,1}^\textrm{I},R_{1,1}^\textrm{I},m^\textrm{I}))
\end{equation}
where $\gamma_{1,1}^\textrm{I}=p_1^\textrm{I}|h_1|^2/\phi\sigma^2$  and $R_{1,1}^\textrm{I}=N_1/m^\textrm{I}$  are the received SINR and the achievable rate of user 1 related to detecting $x_1$  in phase I, respectively. By assuming that $x_1$  is detected when SIC is successful and the fact that in URLLC services, $\varepsilon$  is usually in order of $10^{-5}\sim 10^{-9}$ \cite{survey_1}, the overall decoding error probability at user 1 in phase I can be approximated as
\begin{equation}
    \label{eq.error1I}
    \varepsilon_1^\textrm{I}=\varepsilon_{1,2}^\textrm{I}+(1-\varepsilon_{1,2}^\textrm{I})\varepsilon_{1,1}^\textrm{I}\approx\varepsilon_{1,2}^\textrm{I}+\varepsilon_{1,1}^\textrm{I}.
\end{equation}

Since it is assumed that channels are half-duplex, the relayed signal is not received at user 1. Hence, the overall decoding error probability at user 1 is denoted as $\varepsilon_1=\varepsilon_1^\textrm{I}$ . In contrast, the overall decoding error probability of user 2 depends on user 1 performance and thus the signal of phase II and combining strategy, where the following subsections derive the equations individually for SC and MRC strategies.

\subsection{Selection Combining (SC)}
In this protocol, user 2 does not combine the NOMA phase and relaying phase signals, but decodes transmitted messages from BS and relay (user 1) separately and selects the correctly decoded packet. First, the received message from user 1 in the relaying phase is decoded. If decoding is failed or no signal is received from user 1, then the transmitted message from BS in the NOMA phase is decoded. To differentiate the packets, the packet ID is inserted in the packet head for each device. Therefore, an error occurs when both transmissions are unsuccessful. Decoding error probability of $x_2^\prime$  by user 2 in phase II, i.e., $\varepsilon_{2,2}^\textrm{II}$, is given by
\begin{equation}
    \label{eq.error22II}
     \varepsilon_{2,2}^\textrm{II}\approx Q(f(\gamma_{2,2}^\textrm{II},R_{2,2}^\textrm{II},m^\textrm{II}))
\end{equation}
where $\gamma_{2,2}^\textrm{II}=p_2^\textrm{II}|h_{1,2}|^2/\sigma^2$  and $R_{2,2}^\textrm{II}=N_2/m^\textrm{II}$  are the received SNR and the achievable rate of user 2 related to detecting $x_2^\prime$  in phase II, respectively. One should note that the phase II signal will be transmitted if the message of user 2 is decoded correctly in phase I, so the overall decoding error probability of user 2 in phase II is approximated as
\begin{equation}
    \label{eq.error2II}
    \varepsilon_2^\textrm{II}=\varepsilon_{1,2}^\textrm{I}+(1-\varepsilon_{1,2}^\textrm{I})\varepsilon_{2,2}^\textrm{II}\approx\varepsilon_{1,2}^\textrm{I}+\varepsilon_{2,2}^\textrm{II}.
\end{equation}

Finally, the overall decoding error probability of user 2 in SC strategy is formulated as
\begin{equation}
    \label{eq.errorSC}
    \varepsilon_2=\varepsilon_2^\textrm{I}\varepsilon_2^\textrm{II}\approx\varepsilon_{2,2}^\textrm{I}(\varepsilon_{1,2}^\textrm{I}+\varepsilon_{2,2}^\textrm{II}).
\end{equation}

\subsection{Maximum Ratio Combining (MRC)}
By applying MRC protocol at user 2, since the coding rate of BS-user 2 and user 1-user 2 links are not equal, the determinative link is the bottleneck link, i.e., the link with the lowest coding rate. Therefore, the combined signal with the MRC protocol has a frame of length $m^\text{C}=\max\{m^\text{I},m^\text{II}\}$  symbols and the following SINR
\begin{equation}
    \label{eq.snrMRC}
    \gamma_{2,2}^\text{C}=\frac{m^\text{I}}{m^\text{C}}\gamma_{2,2}^\text{I}+\frac{m^\text{II}}{m^\text{C}}\gamma_{2,2}^\text{II}.
\end{equation}

The probability that user 2 fails in MRC signal decoding is given by
\begin{equation}
    \label{eq.error22MRC}
     \varepsilon_{2,2}^\textrm{C}\approx Q(f(\gamma_{2,2}^\textrm{C},R_{2,2}^\textrm{C},m^\textrm{C}))
\end{equation}
where $R_{2,2}^\textrm{C}=N_2/m^\textrm{C}$  is the achievable rate of user 2 in the combined packet with MRC protocol.

User 2 fails when either its message is decoded correctly by none of them in phase I, or user 1 decodes $x_2$  correctly, but the combined signal is not decoded correctly. Hence, the overall decoding error probability of user 2 in the MRC strategy is given by
\begin{equation}
    \label{eq.errorMRC}
    \varepsilon_2=\varepsilon_{1,2}^\textrm{I}\varepsilon_{2,2}^\textrm{I}+(1-\varepsilon_{1,2}^\textrm{I})\varepsilon_{2,2}^\textrm{C}.
\end{equation}

\section{Problem Formulation}\label{sec.formulation}
In the considered URLLC system, the two users are served with the aim of fairness during two phases with a total $D_{\max}$  symbols period. If channel feedback is available at the transmitter side, users’ data rates can be set according to their instantaneous channel conditions. That being the case, a suitable criterion is max-min fairness \cite{fair_1}.  The throughput of user $i$, $T_i$, is defined as the average bits per each channel use (or complex symbol), which is decoded correctly at the receiver;
\begin{equation}
    \label{eq.throughput}
    T_i\overset{\Delta}{=}\frac{m^\text{I}}{D_{\max}}R_{i,i}^\text{I}(1- \varepsilon_i)
\end{equation}
where $1- \varepsilon_i$  is the reliability of user $i$ and a predefined value for each URLLC use case.

In the C-NOMA scheme, the superposition coding is performed in the NOMA phase, such that the BS enables to transmit users’ signals simultaneously with different powers within a frame of length $m^\text{I}$. User 1 after decoding user 2’s data, sends it in the relaying phase within a frame of length $m^\text{II}$. In Fig. \ref{fig.system}(b) the frame structure of C-NOMA is observed. Therefore, the desired optimization problem is formulated as
\begin{subequations}
\label{eq.opt.prim}
\begin{align}
&\max _{{\left\{ {{p_i^{\rm{I}}},{p_2^{\rm{II}}},{m^j}} \right\}}_{\scriptsize{i = 1,2,} \hfill\atop
			\scriptsize{j = {\rm{I}},{\rm{II}}}\hfill}} \min \left\{ {{T_1},{T_2}} \right\} \label{eq.prim.a} \\
{\rm{ s.t.}}\quad&{m^{\rm{I}}}\left( {p_1^{\rm{I}} + p_2^{\rm{I}}} \right) + {m^{\rm{II}}}p_2^{\rm{II}} \le {D_{\max }}{P_{\rm{ave}}}, \label{eq.prim.b} \\
&0 < p_1^{\rm{I}} + p_2^{\rm{I}} \le {\kappa _{\rm{p}}}{P_{\rm{ave}}},~p_i^{\rm{I}} > 0,~i \in \left\{ {1,2} \right\}, \label{eq.prim.c} \\
&0 \le p_2^{\rm{II}} \le {\kappa _{\rm{p}}}{P_{\rm{ave}}}, \label{eq.prim.d} \\
&{\varepsilon _i} \le {\varepsilon _i}^{\rm{th}},~i \in \left\{ {1,2} \right\}, \label{eq.prim.e} \\
&{m^{\rm{I}}} + {m^{\rm{II}}} = {D_{\max }}.
\label{eq.prim.f}
\end{align}
\end{subequations}

Optimization parameters consist of blocklength and power allocated to two users in phases I and II. Constraint \eqref{eq.prim.b} indicates the system’s total energy consumption budget. Constraints \eqref{eq.prim.c} and \eqref{eq.prim.d} are the general power constraints, where $P_{\text{ave}}$  is the average power, and $\kappa_\text{p}$  is the peak to average power ratio (PAPR) factor. Constraint \eqref{eq.prim.e} guarantees that the decoding error probability of user $i$ does not violate $\varepsilon_i^{\text{th}}$. Moreover, the latency constraint is stated by \eqref{eq.prim.f}.

\section{Problem Solving}\label{sec.solving}
This section will solve the optimization problem in \eqref{eq.opt.prim} for the SC and MRC strategies. To facilitate this issue, we first have to analyze the constraints and specify their optimal status. Let us first consider the constraint \eqref{eq.prim.e} on the acceptable BLER of the two users. Since each URLLC use case needs specific reliability, allocating more resources to achieve a BLER lower than the required  $\varepsilon_i^{\text{th}}$, wastes the rare resources. Moreover, according to \eqref{eq.rate}, the lower desired error probability, the lower data rate. Therefore, $\varepsilon _i = \varepsilon _i^{\rm{th}}$  is an optimal choice. About constraint \eqref{eq.prim.b}, invoking \cite[Proposition 1]{NOMA_4}, the acceptable data rate (i.e., $R>0$ ) in (1), is a monotonically increasing function of the corresponding  SNR/SINR. Using the contradiction method, one can prove that to maximize the throughput, the energy constraint holds with equality \cite{relay_8}, i.e., $m^{\rm{I}}\left( p_1^{\rm{I}} + p_2^{\rm{I}} \right) + m^{\rm{II}}p_2^{\rm{II}} = D_{\max }P_{\rm{ave}}$. In addition, the following proposition indicates the ratio of optimal consumed energy in two transmission phases.
\newtheorem{energy}{Proposition}
\begin{energy}
At the optimal solution, the total consumed energy of the two users in phase I is always greater than the consumed energy in phase II, i.e.,  ${m^{\rm{I}}}P_{\rm{sum}} > {m^{\rm{II}}}p_2^{\rm{II}}$, where  $P_{\rm{sum}}\overset{\Delta}{=}p_1^{\rm{I}} + p_2^{\rm{I}}$. (Refer to Appendix A for proof.)
\end{energy}

Furthermore, invoking \cite[Proposition 2]{NOMA_4}, at the optimum point of Problem \eqref{eq.opt.prim}, throughputs of the two users are equal, i.e., $T_1=T_2$ . Following the above discussion, we provide a solution for the optimization problem in \eqref{eq.opt.prim} with both SC and MRC strategies.

\subsection{Optimal Design of Max-Min Fairness in C-NOMA}
Since at the optimal solution  $T_1=T_2$, equation $R_{2,2}^{\rm{I}} = \frac{1 - \varepsilon _1^{{\rm{th}}}}{1 - \varepsilon _2^{{\rm{th}}}}R_{1,1}^{\rm{I}}$  can be derived from \eqref{eq.throughput}. Moreover, the message of user 2 contains the same number of  bits in both phases, so it can be concluded that  $R_{2,2}^{{\rm{II}}} = \frac{m^{\rm{I}}}{m^{\rm{II}}}R_{2,2}^{\rm{I}}$. Consequently, the optimization problem in \eqref{eq.opt.prim} is rewritten as follows
\begin{subequations}
\label{eq.opt.sec}
\begin{align}
\mathop {\max }\limits_{\left\{ {{m^{\rm{I}}},p_1^{\rm{I}},p_2^{\rm{II}}} \right\}}& {T_1} =
\frac{m^{\rm{I}}}{D_{\max }}\left( {1 - \varepsilon _1^{\rm{th}}} \right)R_{1,1}^{\rm{I}}\label{eq.sec.a}\\
{\rm{ s.t.}}\quad&{m^{\rm{I}}}{P_{\rm{sum}}} + {m^{\rm{II}}}p_2^{\rm{II}} = {D_{\max }}{P_{\rm{ave}}}\label{eq.sec.b}\\
&0 < {P_{\rm{sum}}} \le {\kappa _{\rm{p}}}{P_{\rm{ave}}},~ 0< p_1^{\rm{I}} < \tfrac{P_{\rm{sum}}}{2}\label{eq.sec.c}\\
&0 \le p_2^{{\rm{II}}} \le {\kappa _{\rm{p}}}{P_{\rm{ave}}},~{m^{\rm{I}}}{P_{\rm{sum}}} > {m^{\rm{II}}}p_2^{\rm{II}}\label{eq.sec.d}\\
&{\varepsilon _i} = {\varepsilon _i}^{\rm{th}},~i \in \left\{ {1,2} \right\}\label{eq.sec.e}\\
&{m^{\rm{I}}} + {m^{\rm{II}}} = {D_{\max }}.\label{eq.sec.f}
\end{align}
\end{subequations}
The restriction on $p_1^{\rm{I}}$  in \eqref{eq.sec.c} is applied based on the assumption that  $|h_1|^2>|h_2|^2$. So, to perform SIC correctly in the NOMA phase, it is necessary that  $p_2^\text{I}>p_1^\text{I}$. This problem can be solved using exhaustive linear search; however, we shorten more the search range of $p_1^{\rm{I}}$ to reduce the computational complexity. The main idea can be summarized as follows:
\begin{itemize}
    \item First, by considering user 1’s decoding error probability, i.e., $\varepsilon_1\approx\varepsilon_{1,2}^\textrm{I}+\varepsilon_{1,1}^\textrm{I}$ , the $p_1^{\rm{I}}$  bound that guarantees ${\varepsilon _1} \le {\varepsilon _1}^{\rm{th}}$  is determined. According to our previous work in \cite{NOMA_4},  $\varepsilon_1$ is convex in $p_1^{\rm{I}}$ and at most two values hold the  ${\varepsilon _1}(p_1^{\rm{I}}) = {\varepsilon _1}^{{\rm{th}}}$. With $R_{1,1}^{\rm{I}}=0$  and constant values of $m^{\rm{I}}$  and  $P_{\rm{sum}}$, we obtain the possible solutions that keep this equality in the range of  $0< p_1^{\rm{I}} < \tfrac{P_{\rm{sum}}}{2}$. Clearly, $\varepsilon_{1,1}^\textrm{I}$  is a monotonically decreasing function of  $p_1^{\rm{I}}$, so it is derived that $p_1^{{\rm{I,min}}} = \arg \{ {\varepsilon _1}(p_1^{\rm{I}}) \approx \varepsilon _{1,1}^{\rm{I}}(p_1^{\rm{I}}) = \varepsilon _1^{{\rm{th}}}\}$. On the other hand, $\varepsilon_{1,2}^\textrm{I}$  a monotonically increasing function of $p_1^{\rm{I}}$  yields to  $p_1^{{\rm{I,max}}} = \arg \{ {\varepsilon _1}(p_1^{\rm{I}}) \approx \varepsilon _{1,2}^{\rm{I}}(p_1^{\rm{I}}) = \varepsilon _1^{{\rm{th}}}\}$. Hence, the search region of $p_1^{\rm{I}}$  is given by $p_1^{\rm{I,min}} \le p_1^{\rm{I}} \le p_1^{\rm{I,max}}$.
    \item Since the decoding error probability is a monotonically increasing function of the transmission rate, for each value of  $p_1^{\rm{I}}$  in the feasible range,  $R_{1,1}^{\rm{I}}$ is increased until user 1’s decoding error probability equals to  $\varepsilon _1^{\rm{th}}$. One should note that  $R_{1,1}^{\rm{I}} \le C(\gamma _{1,1}^{\rm{I}})$.
    \item Only those  $p_1^{{\rm{I,min}}} \le p_1^{\rm{I}} \le p_1^{{\rm{I,max}}}$ that satisfy ${\varepsilon _2}(p_1^{\rm{I}}) = {\varepsilon _2}^{{\rm{th}}}$  could be acceptable. Since the decoding error probability of user 2 in both SC and MRC strategies, respectively in \eqref{eq.errorSC} and \eqref{eq.errorMRC}, are increasing function of $p_1^{\rm{I}}$, the transmit power can be obtained using the bisection search method.
    \item After the full search on the values of $m^{\rm{I}}$  and  $P_{\rm{sum}}$, among the feasible solutions, the answer that maximizes $T_1$  is optimal.
\end{itemize}

Based on the above analysis, the algorithm for solving Problem \eqref{eq.opt.sec} is proposed in Algorithm 1. It first determines the local maximum of $T_1$, i.e., ${T_0}^\dag$, by taking constant $m^{\rm{I}}$  and checking all possible values of  $P_{\rm{sum}}$  and $p_1^{\rm{I}}$. In each iteration, the bisection search is adopted to find the desired  $p_1^{\rm{I}}$. By repeating this process on all possible $m^{\rm{I}}$  with a positive integer value, the global maximum of $T_1$, i.e., ${T_0}^ *$, is found. Thus, using a three-dimensional (3-D) exhaustive linear search, the globally optimal solution is achieved.

\begin{algorithm}
\DontPrintSemicolon
 	\textbf{Input}: total blocklength $D_{\max}$, overall BLER of user $i$  ${\varepsilon _i}^{\rm{th}}$, BS average power $P_{\rm{ave}}$, required accuracy $\epsilon$.\\ 
 	\textbf{Output}: optimum power $p{_1^{\rm{I}*}}$, $p{_2^{\rm{I}*}}$,  $p{_2^{\rm{II}*}}$, and blocklength  ${m^{\rm{I}*}}$,  ${m^{\rm{II}*}}$, and fair throughput  ${T_1} = {T_2} = {T_0}^*$.\\
 	\For{${m^{\rm{I}}} = 1:{D_{\max}}$}{
 	\For{${P_{\rm{sum}}} = 0:\Delta p:{\kappa _{\rm{p}}}{P_{\rm{ave}}}$}{
 	Set ${m^{\rm{II}}}: = {D_{\max }} - {m^{\rm{I}}}$ and $p_2^{\rm{II}}: = {{\left( {{D_{\max }}{P_{\rm{ave}}} - {m^{\rm{I}}}{P_{\rm{sum}}}} \right)} / {m^{\rm{II}}}}$.\;
 	\If{$0 \le p_2^{\rm{II}} \le {\kappa _{\rm{p}}}{P_{\rm{ave}}}$ \& ${m^{\rm{I}}}{P_{\rm{sum}}} \ge {m^{\rm{II}}}p_2^{\rm{II}}$}{
 	Calculate $p_1^{{\rm{I}},\min }$ and $p_1^{{\rm{I}},\max }$. \;
 	Set $p_1^{\rm{I}}: = p_1^{\rm{I,min}}$.\;
 	\While{${\varepsilon _2} < \varepsilon _2^{\rm{th}}$}{
 	Set $p_1^{\rm{I}}: = \min \left( {p_1^{\rm{I}} + \Delta p,p_1^{\rm{I,max}}} \right)$.\;
 	Find $R{_{1,1}^{{\rm{I}}^\dag} } = \arg \left\{ {{\varepsilon _1} = \varepsilon _1^{\rm{th}}} \right\}$ via bisection method with accuracy $\epsilon$.\;
 	Calculate $\varepsilon_2$ by (9)/(12)  for SC/MRC.\;
 	}
 	Set $p_1^{{\rm{I,lb}}}: = p_1^{\rm{I}} - \Delta p$ and $p_1^{\rm{I,ub}}: = p_1^{\rm{I}}$.\;
 	Find $p{_1^{{\rm{I}}^\dag} } \in \left[ {p_1^{\rm{I,lb}},p_1^{\rm{I,ub}}} \right]$ that satisfies ${\varepsilon _2} = \varepsilon _2^{\rm{th}}$  via bisection method with accuracy $\epsilon$.\;
 	}
 	}
 	Set $R{_{1,1}^{{\rm{I}}^{\ddag }}}: = \max \left\{ {R{_{1,1}^{{\rm{I}}^\dag}} \left| {{\varepsilon _2} = \varepsilon _2^{\rm{th}}} \right.} \right\}$ and ${T_0}^\dag : = {{\left( {1 - \varepsilon _1^{\rm{th}}} \right){m^{\rm{I}}}R{_{1,1}^{\rm{I}^{\ddag }}}} / {D_{\max }}}$.\;
 	}
 	Set ${T_0}^* : = {\max \{ {{T_0}^\dag } \}}$.\;
 	\textbf{Return} $\left\{ {{m^{\rm{I}*}},p{_1^{\rm{I}*}} ,p{_2^{\rm{II}*}}} \right\} = \arg \max \{ {{T_0}^\dag } \}$, ${m^{\rm{II}*}} = {D_{\max }} - {m^{\rm{I}*}}$, $p{_2^{\rm{I}*}} = \frac{\left( {{D_{\max }}{P_{\rm{ave}}} - {m^{\rm{II}}}^*p{_2^{\rm{II}*}}} \right)}{{m^{\rm{I}}}^*} - p{_1^{\rm{I}*}}$.
    \caption{Optimum Power and Blocklength Allocation Algorithm in the C-NOMA Scheme with SC/MRC Strategy}
 	\label{alg.cnoma}
\end{algorithm}

\subsection{Suboptimal Design of Max-Min Fairness in C-NOMA}
Although the search bounds of the optimum solution of Problem \eqref{eq.opt.sec} stated in Algorithm 1 have been limited, the computational complexity is still high. Now we propose a suboptimal solution to this problem. If phase II transmission is not successful, part of the resources will go to waste, which in turn, will cause the system throughput reduction below the NOMA scheme’s one. Therefore, to avoid this condition and decrease the decoding error probability in phase II, $x_2^\prime$ is transmitted with the maximum power, i.e., $p_2^{\rm{II}} = \kappa _{\rm{p}}{P_{\rm{ave}}}$. Hence, the summation of two users’ transmit power in phase I is calculated as $P_{\rm{sum}} = \left[ \left(D_{\max}P_{\rm{ave}} - m^{\rm{II}}p_2^{\rm{II}} \right)/ m^{\rm{I}} \right]^ + $, where ${\left[ x \right]^ + } \overset{\Delta}{=} \max \left\{ {x,0} \right\}$. Then, as before, the local maximum of $T_1$, i.e.,  ${T_0}^\dag$, is obtained by searching on the possible values of $p_1^{\rm{I}}$  within the range of  $\left[ {p_1^{{\rm{I,}}\min },p_1^{{\rm{I,}}\max }} \right]$. By repeating this process on all possible integer values of $m^{\rm{I}}$  that satisfy  ${m^{\rm{I}}}{P_{{\rm{sum}}}} \ge {m^{\rm{II}}}p_2^{\rm{II}}$, the global maximum of $T_1$, i.e.,  ${T_0}^ *$, is found. If $m^{\rm{I}} = D_{\max }$ , or equivalently  $m^{\rm{II}}=0$, then $P_{\rm{sum}} = P_{\rm{ave}}$. In this case, signal transmission in phase II does not occur, and the C-NOMA scheme is transformed into the NOMA. This suboptimal algorithm which is a special case of Algorithm 1, needs a two-dimensional (2-D) linear search on $\left\{ p_1^{\rm{I}},m^{\rm{I}} \right\}$. The numerical results in section \ref{sec.results} demonstrate that the performance of the suboptimal solution  is slightly worse than the optimal one, while has much lower computational complexity.

\subsection{Computational Complexity}
The computational complexity of Algorithm 1 is calculated as follows. In the first step, to obtain the bounds of  $p_1^{\rm{I}}$, a linear search with complexity $\Omega _1$ is applied. In the next step, $R_{1,1}^{\rm{I}}$  is derived via the bisection method with complexity around ${\log _2}({{\varepsilon _1^{{\rm{th}}}}/\epsilon})$  where $\epsilon$ is the desired accuracy. Besides, the complexity of computing $\varepsilon _2$  is denoted as $\Omega _2$. This step is performed at most ${K_1} = {{(p_1^{{\rm{I,max}}} - p_1^{{\rm{I,min}}})} / {\Delta p}}$  times where ${\Delta p}$  is the search step, so its complexity is denoted as ${K_1}\left({\log _2}({{\varepsilon _1^{{\rm{th}}}}/\epsilon})+\Omega _2\right)$. In the last step, finding $p_1^{\rm{I}}$  via the bisection search method has complexity around  ${\log _2}({{\varepsilon _2^{{\rm{th}}}}/\epsilon})$. These three steps are repeated on the possible values of $P_{\rm{sum}}$  and  $m^{\rm{I}}$, respectively ${K_2} = {{{\kappa _{\rm{p}}}{P_{\rm{ave}}}} / {\Delta p}}$  and $D_{\max }$  times. Therefore, the worst-case complexity of Algorithm 1 is ${\cal O}\left( {{K_2}{D_{\max }}\left({\Omega _1} + {K_1}({\log _2}({{\varepsilon _1^{\rm{th}}}/\epsilon}) + {\Omega _2}) + {\log _2}({{\varepsilon _2^{{\rm{th}}}}/\epsilon})\right)} \right)$. 

Likewise, the computational complexity of the suboptimal algorithm is determined based on the above analysis. However, since $p_2^{\rm{II}}$ is a constant value, $P_{\rm{sum}}$  is removed from the search process. Hence, the worst-case complexity of this algorithm is ${\cal O}\left( {D_{\max }}\left({\Omega _1} + {K_1}({\log _2}({{\varepsilon _1^{{\rm{th}}}}/\epsilon}) + {\Omega _2}) + {\log _2}({{\varepsilon _2^{{\rm{th}}}}/\epsilon})\right) \right)$. 

Although the number of iterations of the proposed algorithms for both SC and MRC techniques is equal, the number of basic operations related to computing the user 2’s decoding error, i.e., $\varepsilon_2$ , is different. According to \eqref{eq.errorSC}, calculation of $\varepsilon_2$ in the SC technique just includes one summation and one multiplication; while, calculation of $\varepsilon_2$ in the MRC technique, regarding \eqref{eq.errorMRC}, requires three summations (one is due to $\gamma_{2,2}^\textrm{C}$) and two multiplications. 

\section{Extension to Multi-User Scenario}\label{sec.extension}
This section considers a more general situation shown in Fig. \ref{fig.system}(a) when there are more than two users in a cell.

\subsection{Problem Formulation}
Let us denote the total number of users as $2K$, and the set of users as  ${\cal K} = \left\{ {1,2, \ldots ,2K} \right\}$. We assume that the users’ channel gains are arranged in descending order, i.e., $| {h_1} |^2 > | {h_2} |^2 >  \cdots  > | {h_{2K}} |^2$.  To implement the NOMA scheme, users are grouped into some clusters. While NOMA distinguishes the users in one cluster, the various clusters become distinct by the OMA technique. Usually, in practice, to decrease the receiver’s complexity, the number of users in each cluster is not considered more than four. Here we form clusters with two users and apply the C-NOMA scheme in each pair. Since for relaying, the two users need to be in the coverage area of each other; pairing is done concerning their relative locations. The number of 2-user clusters is $K$ in the considered network, but the number of possible pairing states is completely random respecting the network topology and is denoted by $Q$. The throughput function of pairing in State $q$, where  $q = 1, \ldots ,Q$, is defined as follows
\begin{equation}
    \label{eq.MU.thrghpt}
    {f_q}\left( {{{\bf{A}}^q},p_i^{\rm{I}},p_j^{\rm{II}},m_{i,j}^{\rm{I}}} \right) = a_{i,j}^q T_0^{i,j};\quad i,j \in {\cal K}.
\end{equation}
Let ${{\bf{A}}^q} = {\left[ {a_{i,j}^q} \right]_{2K \times 2K}}$ be the pairing matrix in State $q$. Here $a_{i,j}^q$ denotes the link between users $i$  and $j$ in State $q$ where
\begin{equation}
    \label{eq.MU.indctr}
    a_{i,j}^q = \left\{ \begin{array}{l}
1,~{\textrm{ if users }}i{\textrm{ and }}j{\textrm{ are paired,}}\\
0,~{\textrm{ otherwise.}}
\end{array} \right.
\end{equation}

The goal is to find the optimum pairing that maximizes the minimum throughput of the cell users. Thus, the optimization problem can be formulated as
\begin{subequations}
    \label{eq.MU.clus}
    \begin{align}
\mathop {\max }\limits_{q = 1, \ldots ,Q}& \mathop {\min }\limits_{{{\bf{A}}^q} = \left[ {a_{i,j}^q} \right]} {f_q}\left( {{{\bf{A}}^q},p_i^{\rm{I}},p_j^{\rm{II}},m_{i,j}^{\rm{I}}} \right)\label{eq.clus.a}\\
{\rm{ s.t.}}\quad& a_{i,j}^q = a_{j,i}^q;~i,j \in {\cal K}\label{eq.clus.b}\\
&\sum\nolimits_{j \in {\cal K}\backslash i} {a_{i,j}^q}  \le 1,~i \in {\cal K}\label{eq.clus.c}\\
&\sum\nolimits_{i \in {\cal K}\backslash j} {a_{i,j}^q}  \le 1,~j \in {\cal K}.\label{eq.clus.d}
    \end{align}
\end{subequations}
Constraint \eqref{eq.clus.b} shows that the pairing matrix ${\bf{A}}^q$ is symmetric. Moreover, constraints \eqref{eq.clus.c} and \eqref{eq.clus.d} indicate that users $i$  and $j$  cannot belong to more than one pair. The inter-programming problem of Problem \eqref{eq.MU.clus} that applies the C-NOMA scheme in each pair is expressed as follows
\begin{subequations}
    \label{eq.MU.RA}
    \begin{align}
T_0^{i,j} =& \mathop {\max }\limits_{\left\{ {m_{i,j}^{\rm{I}},p_i^{\rm{I}},p_j^{{\rm{II}}}} \right\}} \min \left\{ {{T_i},{T_j}} \right\},~\forall a_{i,j}^q = 1 \label{eq.RA.a}\\
{\rm{ s.t.}}\quad& m_{i,j}^{\rm{I}}\left( {p_i^{\rm{I}} + p_j^{\rm{I}}} \right) + m_{i,j}^{\rm{II}}p_j^{\rm{II}} =  \tfrac{{D_{\max }}{P_{\rm{ave}}}}{K} \label{eq.RA.b}\\ 
&0 < p_i^{\rm{I}} + p_j^{\rm{I}} \le {\kappa _{\rm{p}}}{P_{\rm{ave}}} \label{eq.RA.c}\\
&0 \le p_j^{\rm{II}} \le {\kappa _{\rm{p}}}{P_{\rm{ave}}} \label{eq.RA.d}\\
&{\varepsilon _i} = {\varepsilon _i}^{\rm{th}},~{\varepsilon _j} = {\varepsilon _j}^{\rm{th}} \label{eq.RA.e}\\
&m_{i,j}^{\rm{I}} + m_{i,j}^{\rm{II}} = {D_{\max }}. \label{eq.RA.f}
    \end{align}
\end{subequations}

Here it is assumed that  $| {h_i} |^2 > | {h_j} |^2$ so $p_i^{\rm{I}} < p_j^{\rm{I}}$. Constraint \eqref{eq.RA.b} indicates that the total system’s energy consumption is distributed equally among the pairs. For solving Problem \eqref{eq.MU.clus}, it is needed that problem \eqref{eq.MU.RA} is solved for all the potential pair-users in State  $q$, i.e.,  $\forall a_{i,j}^q = 1;~i,j = 1, \ldots ,2K$. Hence, to find the optimum pairing, the inter-programming problem has to be solved $QK$  times. By an exhaustive search over all the neighboring users, every two users are paired that the minimum achieved throughput in the cell is maximized. The complexity of the exhaustive search (i.e., the number of iterations needed to find the optimal pairing) is almost high, resulting in excessive scheduling delay with a large number of users. To alleviate the computational complexity, a suboptimal pairing algorithm is proposed in the following subsection.

\subsection{The proposed C-NOMA pairing}
Here, a suboptimal solution for Problem \eqref{eq.MU.clus} is proposed. The objective is to maximize the throughput of the weakest user among all $2K$ users by allocating them into different pairs according to the geographic locations. To implement the proposed user pairing, the graph matrix of the network topology has to be obtained first. For this purpose, each user ought to find all the users in its coverage area with radius $r_0$. Since the aim is leveraging C-NOMA to increase reliability and system capacity, the priority is with C-NOMA pairs starting from the weakest user. Users that are far from others and do not have a chance to exploit the C-NOMA technique use NOMA or OMA instead, depending on their channel condition. Finally, the users that have not been scheduled in C-NOMA pairs are rearranged to form hybrid NOMA/OMA pairs, as will be described in the following subsection. Algorithm 2 expresses the proposed C-NOMA-based user pairing in detail. The fact that how frequently the pairing process is executed mainly depends on the URLLC use case. For example, in factory automation with fixed or slow speed devices, the algorithm does not need to perform in each frame. Moreover, it should be noted that these computations are performed at the BS with the assumption of CSIT, and the results are sent to the users. 

\begin{algorithm}
\DontPrintSemicolon
     \textbf{Input}: sorted DL channel gains in descending order ${| {h_1} |^2} > {| {h_2} |^2} >  \cdots  > {| {h_{2K}} |^2}$ and the corresponding D2D channel gains, device coverage radius $r_0$, inputs of Algorithm \ref{alg.cnoma}.\\
     \textbf{Output}: the user pairing ${\bf{A}} = {\left[ {a_{i,j}} \right]_{2K \times 2K}}$. \\
     Determine the graph matrix of the network topology.\;
     Set $i: = 2K$.\;
     \While{$i \ge 1$}{\tcp*{allocating C-NOMA pairs}
     \If{user $i$ has not been paired }{
     Find the set of unpaired adjacent users of user $i$, ${{\bm{\psi }}_i}$.\;
     \If{${\rm{length}}\left( {{\bm{\psi }}_i} \right) \ne 0$}{
     \For{$l = 1:{\rm{length}}\left( {{\bm{\psi }}_i} \right)$}{
     Calculate $T_0^{i,{{\bm{\psi }}_i}(l)}$ by Algorithm 1.\;
     }
     Set $\left[ {T_0^*,index} \right]: = \max \left\{ {T_0^{i,{{\bm{\psi }}_i}(l)}} \right\}$ and $j: = {{\bm{\psi }}_i}(index)$.\;
     Pair users $i$ and $j$, i.e. ${a_{i,j}}: = 1$.\;
     }
     }
     Set $i: = i - 1$.\;
     }
     Set $i: = 2K$ and $j: = 1$.\;
     \While{$i > j$}{\tcp*{allocating hybrid pairs}
     \If{user $i$ has not been paired }{
     \While{user $j$ has been paired }{
     Set $j: = j + 1$.\;
     }
     Pair users $i$ and $j$, i.e. ${a_{i,j}}: = 1$.\;
     Set $j: = j + 1$.\;
     }
     Set $i: = i - 1$.\;
     }
     \textbf{Return}: ${\bf{A}} = {\left[ {a_{i,j}} \right]_{2K \times 2K}}$.
     \caption{Joint suboptimal C-NOMA-based user pairing and resource allocation}
 	 \label{alg.clustering}
\end{algorithm}

\subsection{Hybrid pairing}
\begin{figure}
    \centering
    \includegraphics[width=0.8\columnwidth,trim={6cm 18.7cm 8cm 2.3cm},clip]{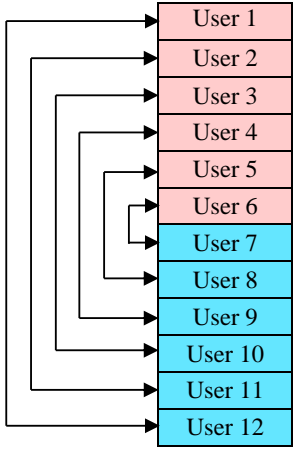}
    \caption{The 2-user NOMA pairing scheme \cite{cluster_1}.}
    \label{fig.NOMA.clustr}
\end{figure}

To describe the hybrid pairing, let us first consider the  NOMA user pairing scheme proposed in \cite{cluster_1}. Pursuant to this, the first strong user is paired with the first weak user; the second strong user is paired with the second weak user, and so on. Accordingly, all the users are paired. The fact is that principle of NOMA is to select users with a high difference in their channel gains. In particular, NOMA’s performance diminishes when the difference in channel gains among the users is small. For example, in Fig. \ref{fig.NOMA.clustr}, user-pairs 6 and 7, which have almost the same channel conditions, may decrease the spectral efficiency and system capacity due to the unsuccessful SIC. Hence, it is sensible that such non-suitable pairs are omitted from NOMA scheduling, and their clustering continues with OMA. In this method, the BS adaptively switches between the NOMA and OMA transmission modes according to the instantaneous strength of wireless channels and hence the performance of the NOMA/OMA user pairing, and each of them that meets the max-min fairness criterion is selected as the access scheme.

We discussed the hybrid pairing for scheduling the users that are isolated or left unpaired in the proposed C-NOMA-based user pairing. However, these two basic schemes, namely NOMA and hybrid user pairing, can independently be implemented and are considered as benchmark schemes in our simulations.

\section{Numerical Results}\label{sec.results}
In this section, the proposed C-NOMA scheme’s performance along with SC and MRC strategies are evaluated through the numerical results based on our analytical solutions. A heterogeneous network consists of URLLC users with different reliability requirements is considered. PAPR factor and required accuracy in Algorithm \ref{alg.cnoma} are considered as $\kappa _{\rm{p}} = 1.2$  and  $\epsilon = 10^{ - 15}$, respectively. Also, it is assumed that $P_{\rm{ave}} = 10{\rm{~W}}$  and $D_{\max } = 200$  channel uses, unless otherwise stated. The numerical results are provided based on fixed channel gains with two users and random channel gains with more than two users, which are presented in the following two subsections.

\subsection{Two-user Network with Fixed Channel Gains}
Throughout this subsection, to provide insight into the relationships between the proposed and the benchmark schemes, the channel gains of the two users are set to be fixed. For instance, it is assumed that $|{h_1}|^2/{\sigma ^2} = 0.8$  and $|{h_2}|^2/{\sigma ^2} = 0.1$. We investigate the performance of the proposed schemes in two various relaying link status. Meaning, when the two users are near to each other and the relaying link is strong, it is assumed that $|{h_{1,2}}|^2/{\sigma ^2} = 0.5$, and when the two users are far from each other, and the relaying link is poor, it is assumed that $|{h_{1,2}}|^2/{\sigma ^2} = 0.01$. Meanwhile, users BLER are considered as $\varepsilon _1^{\rm{th}} = 10^{ - 7}$  and  $\varepsilon _2^{\rm{th}} = 10^{ - 5}$.

\begin{figure}
    \centering
    \includegraphics[width=1\columnwidth,trim={4cm 8.4cm 4cm 8.4cm},clip]{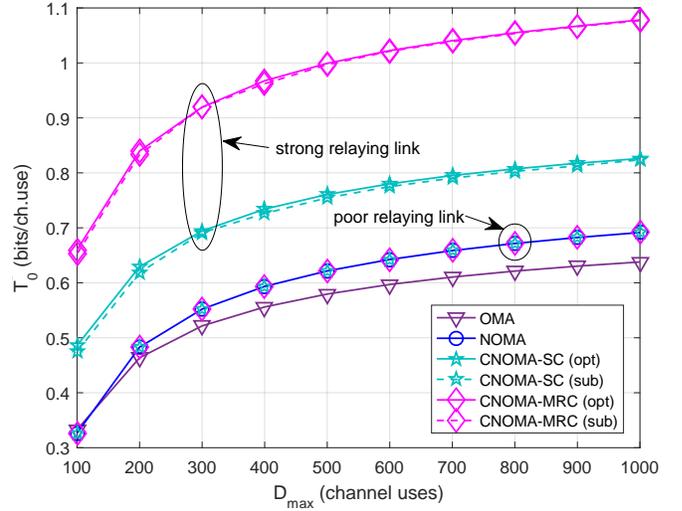}
    \caption{Maximum fair throughput achieved by the C-NOMA and NOMA schemes versus $D_{\max}$, when $P_{\rm{ave}} = 10{\rm{~W}}$. }
    \label{fig.BL_var}
\end{figure}

In Fig. \ref{fig.BL_var}, the effect of total blocklength, $D_{\max}$, on the fair throughput in the proposed C-NOMA with SC and MRC strategies is assessed in two relaying link modes. Also, the optimal NOMA and OMA results in our previous work \cite{NOMA_4} are shown for comparison. It is observed that in the strong relaying link mode, both combining strategies applied to the C-NOMA effectively improve the fair throughput compared to the NOMA/OMA. It is also observed that the MRC receiver outperforms the SC receiver, regardless of the blocklength. Because in the combined signal with MRC protocol, SINR increases, so the decoding error probability of user 2 decreases. Hence, it is possible that by less blocklength allocation to phase II, the reliability performance of user 2 can still be guaranteed at the desired level. As a result, more blocklength is allocated to phase I. Hence, users’ data rates and system fair throughput increase.

On the other hand, in a poor relaying link, the C-NOMA scheme (in both combining strategies) has exactly the same performance as the NOMA. In fact, in this case, the optimal decision is in favor of the direct link, and the C-NOMA is transformed into the NOMA. However, in a realistic wireless channel, mixed conditions occur together, and C-NOMA outperforms the NOMA on average. Moreover, it is observed that suboptimal solutions in both SC and MRC receivers converge to the near-optimal solutions. 

\begin{figure}
    \centering
    \includegraphics[width=1\columnwidth,trim={4cm 8.4cm 4cm 8.4cm},clip]{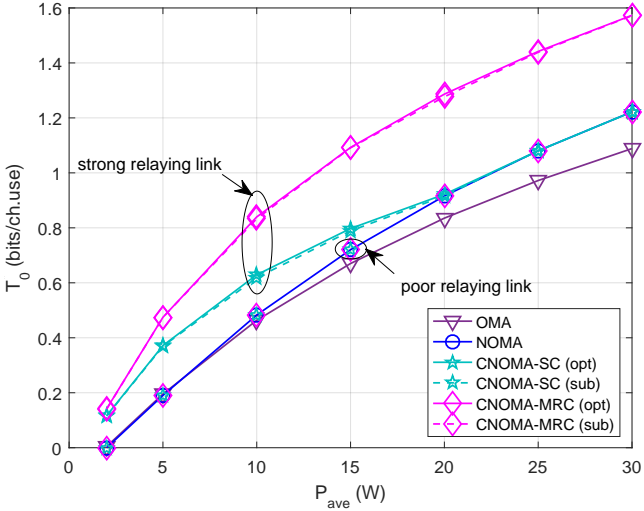}
    \caption{Maximum fair throughput achieved by the C-NOMA and NOMA schemes versus  $P_{\rm{ave}}$, when $D_{\max} = 200$. }
    \label{fig.pwr_var}
\end{figure}

In Fig. \ref{fig.pwr_var}, the effect of average total power, $P_\textrm{ave}$, on the fair throughput is investigated. In the strong relaying link mode, the C-NOMA’s superiority with MRC receiver is notable against the SC receiver and the NOMA/OMA scheme. In addition, the C-NOMA with SC strategy outperforms the NOMA in low power/SNR ranges, while it coincides with the NOMA on average powers greater than 20 W. This could be justified by the fact that in SC strategy, the signals do not combine, and transmission in phase II assures the success of user 2’s packet decoding. Hence, in low SNRs where the weak user’s probability of successful decoding in phase I is not too high, the reliability is increased by retransmission in phase II. However, in high SNRs, where the allocated power of user 2 in the NOMA phase guarantees the reliability, phase II transmission is pointless. Therefore, in this case, transmission via a single phase is optimal in comparison with two-phase, and the proposed scheme performs like the NOMA. Moreover, in the poor relaying link mode, the C-NOMA scheme always complies with the NOMA. As a result, from the complexity perspective, the C-NOMA usage with SC strategy seems sensible just in low SNR regimes.

\subsection{Multi-user Network with Random Channel Gains}
Here, we assume that the BS is located at the center of a cell with radius of $300$ m. The system bandwidth is set as $B = 1$ MHz, which is equivalent to a DL transmission duration $0.2$ ms for a blocklength of $200$ channel uses, and satisfies the low-latency criterion of URLLC standards. The noise power spectral density is $-173$ dBm/Hz, and small-scale channel coefficients are Rayleigh fading with ${\cal C}{\cal N}\left( {0,1} \right)$ distribution. Large-scale path loss is modeled as $L = 35.3 + 37.6{\log _{10}}d({\rm{m}})$ dB \cite{relay_8}. The total number of independent channel generations is set as $1000$.

\begin{figure}
    \centering
    \includegraphics[width=1\columnwidth,trim={4cm 8.4cm 4cm 8.4cm},clip]{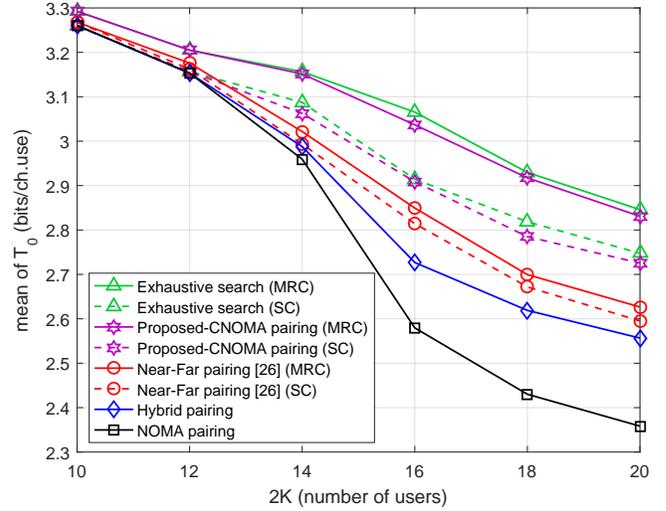}
    \caption{Average fair throughput achieved by the different pairing schemes versus the number of users.}
    \label{fig.mu_thrpt}
\end{figure}

Fig. \ref{fig.mu_thrpt} illustrates the average achievable fair throughput versus the number of users, $2K$ , for the proposed C-NOMA pairing with SC and MRC strategies. We compare it with exhaustive search method and the method proposed in \cite{cluster_3}, which is based on pairing one near user and one far user. In that method, first we sort the $K^2$ D2D channel gains of any near-far pair and delete the $K-1$ weakest channels. Then, by considering the number of deleted channel gains of each far user, first a near user is paired to the far user with the largest-number deleted channel and last the far user with the smallest-number. The selection criteria is to maximize the throughput of the far user. To be comparable with our proposed method, unlike \cite{cluster_3}, we assume that both near and far users can communicate with the BS directly, and SC/MRC combining schemes are performed at the far user. Moreover, the NOMA and hybrid pairing schemes are illustrated as benchmark. It demonstrates that the proposed C-NOMA pairing scheme (in both MRC and SC techniques) converges to a near-optimal solution. While, the near-far pairing method achieves the lower performance in both combining strategies. On the other hand, the NOMA pairing scheme stated in \cite{cluster_1} yields the lowest throughput, especially in the presence of a large number of users, and as expected, the hybrid pairing scheme outperforms the NOMA pairing.

\begin{figure}
    \centering
    \includegraphics[width=1\columnwidth,trim={4cm 8.4cm 4cm 8.4cm},clip]{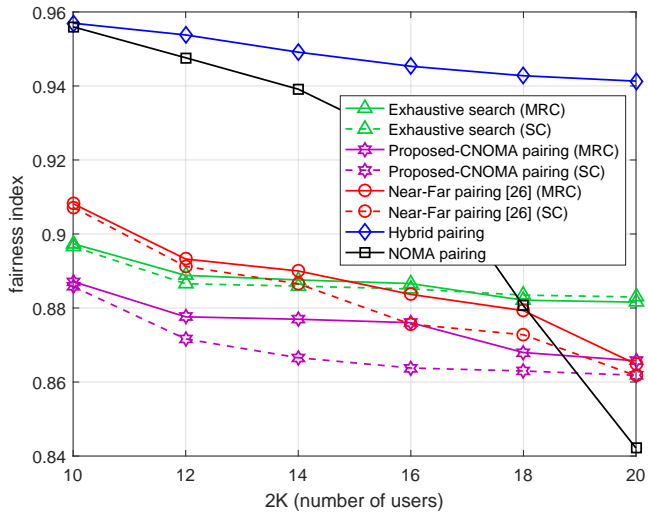}
    \caption{Fairness comparison between the different pairing schemes versus the number of users.}
    \label{fig.mu_fair}
\end{figure}

To evaluate the fairness of the proposed C-NOMA-based user pairing, Fig. \ref{fig.mu_fair} indicates Jain’s fairness index for the proposed scheme and the benchmarks. Jain’s fairness index is defined as \cite{fair_2}
\begin{equation}
    J = \frac{{\left( {\sum\nolimits_{k = 1}^K {T_k^*} } \right)}^2}{K\sum\nolimits_{k = 1}^K {T{{_k^*}^2}} },
    \label{eq.jain}
\end{equation}
where $T_k^*$  indicates the optimal fair throughput of pair $k$. Jain’s fairness index is bounded in $[0,1]$  which equal users’ throughput obtains the maximum value. As Fig. \ref{fig.mu_fair} illustrates, the hybrid pairing scheme is fairer comparing to the C-NOMA-based and the NOMA pairing schemes. The reason is that in the C-NOMA-based pairing schemes, i.e., the proposed, near-far, and exhaustive search methods, creating C-NOMA pairs for all the users is not probable. Hence, unavoidably, some users are scheduled in hybrid NOMA/OMA pairs. Since the C-NOMA users will achieve more throughput than the users with hybrid pairing, the fairness will degrade in these schemes.

Moreover, regarding the logic behind the hybrid pairing, it will always be fairer than the NOMA pairing. Interestingly, the C-NOMA-based pairing schemes (with both combining strategies) result in more fairness relative to the NOMA pairing in the presence of a large number of users. This is due to the fact that the denser the network is, the more users will experience the same channel. This will cause more failures in NOMA scheduling, so the C-NOMA pairing will obtain more fairness in that case.

\section{Conclusion and Future Works}\label{sec.conclusion}
In this paper, the combination of NOMA with the cooperative relaying technique (i.e., C-NOMA) was considered in short packet communications to guarantee high reliability and low latency. The performance of two relaying strategies, i.e., SC and MRC, was presented in terms of decoding error probability in a quasi-static channel. Besides, the necessity to provide QoS of all users with critical services motived us to consider max-min fairness as a design criterion in URLLC systems. To this end, first, an optimization problem was formulated for a two-user DL C-NOMA system, and optimal power, blocklength, and transmission rate were determined under the total energy consumption, reliability, and delay constraints. To decrease the computational complexity, a suboptimal algorithm was proposed with near-optimal performance. Numerical results showed that the proposed C-NOMA scheme improves the users’ fair throughput significantly, compared to the NOMA scheme. Moreover, it was demonstrated that the C-NOMA scheme with MRC strategy outperforms SC strategy.

Finally, the problem was extended to a multi-user scenario, and a pairing scheme based on C-NOMA was proposed. Monte Carlo simulations showed that the proposed C-NOMA pairing scheme performs close to the optimal solution, with less computational complexity. Further, the simulation results verify the supremacy of the proposed user pairing (with both SC and MRC techniques) over the near-far pairing method proposed in \cite{cluster_3}, as well as the NOMA and hybrid OMA/NOMA pairing schemes in boost the average fair throughput despite degrading the fairness index slightly.

The presented work in this paper can be extended from different directions. Two of the main potential extensions that remain for future works are developing the proposed pairing algorithm for the case of statistical CSI, and considering a distributed method for network coordination. The statistical CSI knowledge can remove the shortage of the out-dated CSI and the feedback overhead due to CSIT. Despite distributed method for network coordination might look to be more suitable for URLLC, it requires deep investigation as they normally introduce different types of overhead which may cause additional delay. 

\appendices
\section{Proof of Proposition 1}
We prove Proposition 1 by the contradiction method. We consider the optimal solution of problem \eqref{eq.opt.sec} as $\left\{ {p{{_1^{\rm{I}\dag}} },p{{_2^{\rm{I}\dag}} },p{{_2^{\rm{II}\dag}} },{m^{\rm{I}\dag}} ,{m^{\rm{II}\dag}} } \right\}$, where ${m^{\rm{I}\dag}} ({p_1^{{\rm{I}}\dag } + p_2^{{\rm{I}}\dag }}) < {m^{\rm{II}\dag}} p_2^{{\rm{II}}\dag }$. It can achieve the maximum value of $\min \left\{ {{T_1},{T_2}} \right\}$, which is denoted by $T_0^\dag$. We increase $p_1^{{\rm{I}}\dag }$ and $p_2^{{\rm{I}}\dag }$  by multiplying in a scalar value $\alpha  > 1$  to attain ${p_1^{\rm{I}}}^* = \alpha {p_1^{\rm{I}} }^\dag$  and  ${p_2^{\rm{I}}}^* = \alpha {p_2^{\rm{I}} }^\dag$. It can be verified that the following equation holds,
\begin{equation*}
\begin{array}{c}
{m^{\rm{I}\dag}} \left( {p{_1^{\rm{I}\dag }} + p{_2^{\rm{I}\dag}}} \right) + {m^{\rm{II}\dag}} p{_2^{\rm{II}\dag}} = \\
{m^{\rm{I}\dag}} \left( {{p_1^{\rm{I}}}^* + {p_2^{\rm{I}}}^*} \right) + {m^{\rm{II}\dag}} {p_2^{\rm{II}}}^* = {D_{\max }}{P_{\rm{ave}}}
\end{array}
\end{equation*}

We note that since  $\alpha  > 1$, so ${p_1^{\rm{I}}}^* > {p_1^{\rm{I}} }^\dag$  and ${p_2^{\rm{I}}}^* > {p_2^{\rm{I}}}^\dag$. Hence, we have ${\gamma _{1,1}^{\rm{I}}}^* > {\gamma _{1,1}^{\rm{I}}}^\dag$  and
\begin{align*}
{\gamma _{2,2}^{\rm{I}}}^* &= \frac{{{p_2^{\rm{I}}}^*{{\left| {h_2} \right|}^2}}}{{p_1^{\rm{I}}}^*{{\left| {h_2} \right|}^2} + {\sigma ^2}} = \frac{p_2^{{\rm{I}}\dag }{{\left| {h_2} \right|}^2}}{p_1^{{\rm{I}}\dag }{{\left| {h_2} \right|}^2} + \frac{\sigma ^2}{\alpha }}\\ 
&> \frac{p_2^{{\rm{I}}\dag }{{\left| {h_2} \right|}^2}}{p_1^{{\rm{I}}\dag }{{\left| {h_2} \right|}^2} + {\sigma ^2}} = {\gamma _{2,2}^{\rm{I}}}^\dag 
\end{align*}    

This means as ${p_i}^{{\rm{I}}\dag }$,  $i \in \left\{ {1,2} \right\}$, increases to ${p_i^{{\rm{I}}}}^* $, the corresponding SNR/SINR increases, which results in an increase in $R_{i,i}^{\rm{I}}$ and finally $T_i$ increases (Invoking  \cite[Appendix A]{NOMA_4}, the allowed $R_{i,i}^{\rm{I}}$ is a monotonically increasing function of $\gamma _{i,i}^{\rm{I}}$.). On the other hand, $R_{i,i}^{\rm{I}}$ and so $T_i$  are clearly increasing functions of ${m^{\rm{I}}}$. Then, we can construct a new solution $\left\{ {p_1^{\rm{I}}}^*, {p_2^{\rm{I}}}^*,{p_2^{\rm{II}}}^*,{m^{\rm{I}}}^* ,{m^{\rm{II}}}^* \right\}$, where corresponds to $T_0^*$. Also, ${m^{\rm{I}}}^* = {m^{{\rm{I}}\dag }} + \Delta m$ and ${m^{\rm{II}}}^* = {m^{{\rm{II}}\dag }} - \Delta m$ with $\Delta m > 0$. As before, this solution satisfies ${m^{\rm{I}}}^* \left( {{p_1^{\rm{I}}}^* + {p_2^{\rm{I}}}^*} \right) + {m^{\rm{II}}}^* {p_2^{\rm{II}}}^* = {D_{\max }}{P_{\rm{ave}}}$. Since ${p_i^{\rm{I}}}^* > p_i^{{\rm{I}}\dag }$ and ${m^{\rm{I}}}^* > {m^{{\rm{I}}\dag }}$, we have $T_0^* > T_0^\dag$. This contradicts the assumption that $T_0^\dag$ is an optimal solution. So, we can always find a proper $\alpha$ and $\Delta m$  such that ${m^{\rm{I}}}^*\left( {{p_1^{\rm{I}}}^* + {p_2^{\rm{I}}}^*} \right) > {m^{\rm{II}}}^*{p_2^{\rm{II}}}^*$.


%\section*{Acknowledgment}


\ifCLASSOPTIONcaptionsoff
  \newpage
\fi

\bibliographystyle{IEEEtran}
\bibliography{references}

\begin{thebibliography}{10}
\providecommand{\url}[1]{#1}
\csname url@samestyle\endcsname
\providecommand{\newblock}{\relax}
\providecommand{\bibinfo}[2]{#2}
\providecommand{\BIBentrySTDinterwordspacing}{\spaceskip=0pt\relax}
\providecommand{\BIBentryALTinterwordstretchfactor}{4}
\providecommand{\BIBentryALTinterwordspacing}{\spaceskip=\fontdimen2\font plus
\BIBentryALTinterwordstretchfactor\fontdimen3\font minus
  \fontdimen4\font\relax}
\providecommand{\BIBforeignlanguage}[2]{{%
\expandafter\ifx\csname l@#1\endcsname\relax
\typeout{** WARNING: IEEEtran.bst: No hyphenation pattern has been}%
\typeout{** loaded for the language `#1'. Using the pattern for}%
\typeout{** the default language instead.}%
\else
\language=\csname l@#1\endcsname
\fi
#2}}
\providecommand{\BIBdecl}{\relax}
\BIBdecl

\bibitem{eucnc_conf}
F.~Salehi, N.~Neda, M.-H. Majidi, and H.~Ahmadi, ``Max-min fairness with
  selection combining strategy on cooperative noma: A finite blocklength
  analysis,'' in \emph{2021 Joint European Conference on Networks and
  Communications 6G Summit (EuCNC/6G Summit)}, 2021, pp. 43--48.

\bibitem{IT_1}
G.~{Durisi}, T.~{Koch}, and P.~{Popovski}, ``Toward massive, ultrareliable, and
  low-latency wireless communication with short packets,'' \emph{Proceedings of
  the IEEE}, vol. 104, no.~9, pp. 1711--1726, 2016.

\bibitem{IT_2}
Y.~{Polyanskiy}, H.~V. {Poor}, and S.~{Verdu}, ``Channel coding rate in the
  finite blocklength regime,'' \emph{IEEE Transactions on Information Theory},
  vol.~56, no.~5, pp. 2307--2359, 2010.

\bibitem{IT_3}
W.~{Yang}, G.~{Durisi}, T.~{Koch}, and Y.~{Polyanskiy}, ``Quasi-static
  multiple-antenna fading channels at finite blocklength,'' \emph{IEEE
  Transactions on Information Theory}, vol.~60, no.~7, pp. 4232--4265, 2014.

\bibitem{IT_4}
P.~{Wu} and N.~{Jindal}, ``Coding versus {ARQ} in fading channels: How reliable
  should the phy be?'' \emph{IEEE Transactions on Communications}, vol.~59,
  no.~12, pp. 3363--3374, 2011.

\bibitem{IT_5}
B.~{Makki}, T.~{Svensson}, and M.~{Zorzi}, ``Finite block-length analysis of
  the incremental redundancy harq,'' \emph{IEEE Wireless Communications
  Letters}, vol.~3, no.~5, pp. 529--532, 2014.

\bibitem{IT_6}
S.~{Xu}, T.~{Chang}, S.~{Lin}, C.~{Shen}, and G.~{Zhu}, ``Energy-efficient
  packet scheduling with finite blocklength codes: Convexity analysis and
  efficient algorithms,'' \emph{IEEE Transactions on Wireless Communications},
  vol.~15, no.~8, pp. 5527--5540, 2016.

\bibitem{IT_7}
H.~{Ren}, C.~{Pan}, Y.~{Deng}, M.~{Elkashlan}, and A.~{Nallanathan}, ``Joint
  pilot and payload power allocation for massive-mimo-enabled urllc iiot
  networks,'' \emph{IEEE Journal on Selected Areas in Communications}, vol.~38,
  no.~5, pp. 816--830, 2020.

\bibitem{IT_8}
------, ``Resource allocation for secure urllc in mission-critical iot
  scenarios,'' \emph{IEEE Transactions on Communications}, vol.~68, no.~9, pp.
  5793--5807, 2020.

\bibitem{IT_9}
C.~{She}, Y.~{Duan}, G.~{Zhao}, T.~Q.~S. {Quek}, Y.~{Li}, and B.~{Vucetic},
  ``Cross-layer design for mission-critical iot in mobile edge computing
  systems,'' \emph{IEEE Internet of Things Journal}, vol.~6, no.~6, pp.
  9360--9374, 2019.

\bibitem{NOMA_1}
Y.~{Yu}, H.~{Chen}, Y.~{Li}, Z.~{Ding}, and B.~{Vucetic}, ``On the performance
  of non-orthogonal multiple access in short-packet communications,''
  \emph{IEEE Communications Letters}, vol.~22, no.~3, pp. 590--593, 2018.

\bibitem{NOMA_2}
X.~{Sun}, S.~{Yan}, N.~{Yang}, Z.~{Ding}, C.~{Shen}, and Z.~{Zhong},
  ``Short-packet downlink transmission with non-orthogonal multiple access,''
  \emph{IEEE Transactions on Wireless Communications}, vol.~17, no.~7, pp.
  4550--4564, 2018.

\bibitem{NOMA_3}
Y.~{Xu}, C.~{Shen}, T.~{Chang}, S.~{Lin}, Y.~{Zhao}, and G.~{Zhu},
  ``Transmission energy minimization for heterogeneous low-latency noma
  downlink,'' \emph{IEEE Transactions on Wireless Communications}, vol.~19,
  no.~2, pp. 1054--1069, 2020.

\bibitem{NOMA_4}
\BIBentryALTinterwordspacing
F.~Salehi, N.~Neda, and M.-H. Majidi, ``Max-min fairness in downlink
  non-orthogonal multiple access with short packet communications,'' \emph{AEU
  - International Journal of Electronics and Communications}, vol. 114, p.
  153028, 2020. [Online]. Available:
  \url{https://www.sciencedirect.com/science/article/pii/S1434841119321570}
\BIBentrySTDinterwordspacing

\bibitem{relay_1}
Y.~{Hu}, J.~{Gross}, and A.~{Schmeink}, ``On the performance advantage of
  relaying under the finite blocklength regime,'' \emph{IEEE Communications
  Letters}, vol.~19, no.~5, pp. 779--782, 2015.

\bibitem{relay_2}
Y.~{Hu}, A.~{Schmeink}, and J.~{Gross}, ``Blocklength-limited performance of
  relaying under quasi-static rayleigh channels,'' \emph{IEEE Transactions on
  Wireless Communications}, vol.~15, no.~7, pp. 4548--4558, 2016.

\bibitem{relay_3}
------, ``Optimal scheduling of reliability-constrained relaying system under
  outdated csi in the finite blocklength regime,'' \emph{IEEE Transactions on
  Vehicular Technology}, vol.~67, no.~7, pp. 6146--6155, 2018.

\bibitem{C-NOMA_1}
Z.~{Ding}, M.~{Peng}, and H.~V. {Poor}, ``Cooperative non-orthogonal multiple
  access in 5g systems,'' \emph{IEEE Communications Letters}, vol.~19, no.~8,
  pp. 1462--1465, 2015.

\bibitem{C-NOMA_2}
Z.~{Ding}, H.~{Dai}, and H.~V. {Poor}, ``Relay selection for cooperative
  noma,'' \emph{IEEE Wireless Communications Letters}, vol.~5, no.~4, pp.
  416--419, 2016.

\bibitem{C-NOMA_3}
P.~{Xu}, Y.~{Wang}, G.~{Chen}, G.~{Pan}, and Z.~{Ding}, ``Design and evaluation
  of buffer-aided cooperative noma with direct transmission in iot,''
  \emph{IEEE Internet of Things Journal}, pp. 1--1, 2020.

\bibitem{C-NOMA_4}
F.~{Kara} and H.~{Kaya}, ``Threshold-based selective cooperative-noma,''
  \emph{IEEE Communications Letters}, vol.~23, no.~7, pp. 1263--1266, 2019.

\bibitem{C-NOMA_5}
X.~{Lai}, Q.~{Zhang}, and J.~{Qin}, ``Cooperative noma short-packet
  communications in flat rayleigh fading channels,'' \emph{IEEE Transactions on
  Vehicular Technology}, vol.~68, no.~6, pp. 6182--6186, 2019.

\bibitem{relay_7}
Y.~{Hu}, M.~C. {Gursoy}, and A.~{Schmeink}, ``Efficient transmission schemes
  for low-latency networks: Noma vs. relaying,'' in \emph{2017 IEEE 28th Annual
  International Symposium on Personal, Indoor, and Mobile Radio Communications
  (PIMRC)}, 2017, pp. 1--6.

\bibitem{relay_8}
H.~{Ren}, C.~{Pan}, Y.~{Deng}, M.~{Elkashlan}, and A.~{Nallanathan}, ``Joint
  power and blocklength optimization for urllc in a factory automation
  scenario,'' \emph{IEEE Transactions on Wireless Communications}, vol.~19,
  no.~3, pp. 1786--1801, 2020.

\bibitem{cluster_2}
L.~{Zhu}, J.~{Zhang}, Z.~{Xiao}, X.~{Cao}, and D.~O. {Wu}, ``Optimal user
  pairing for downlink non-orthogonal multiple access ({NOMA}),'' \emph{IEEE
  Wireless Communications Letters}, vol.~8, no.~2, pp. 328--331, 2019.

\bibitem{cluster_3}
Y.~{Cheng}, K.~H. {Li}, K.~C. {Teh}, S.~{Luo}, and W.~{Wang}, ``Two-step user
  pairing for {OFDM}-based cooperative {NOMA} systems,'' \emph{IEEE
  Communications Letters}, vol.~24, no.~4, pp. 903--906, 2020.

\bibitem{cluster_4}
J.~{Zhang}, X.~{Tao}, H.~{Wu}, and X.~{Zhang}, ``Performance analysis of user
  pairing in cooperative {NOMA} networks,'' \emph{IEEE Access}, vol.~6, pp.
  74\,288--74\,302, 2018.

\bibitem{cluster_5}
P.~{Hũu}, M.~A. {Arfaoui}, S.~{Sharafeddine}, C.~M. {Assi}, and A.~{Ghrayeb},
  ``A low-complexity framework for joint user pairing and power control for
  cooperative {NOMA} in 5g and beyond cellular networks,'' \emph{IEEE
  Transactions on Communications}, vol.~68, no.~11, pp. 6737--6749, 2020.

\bibitem{cluster_6}
A.~K. {Lamba}, R.~{Kumar}, and S.~{Sharma}, ``Joint user pairing, subchannel
  assignment and power allocation in cooperative non-orthogonal multiple access
  networks,'' \emph{IEEE Transactions on Vehicular Technology}, vol.~69,
  no.~10, pp. 11\,790--11\,799, 2020.

\bibitem{IT_10}
C.~{She}, C.~{Yang}, and T.~Q.~S. {Quek}, ``Joint uplink and downlink resource
  configuration for ultra-reliable and low-latency communications,'' \emph{IEEE
  Transactions on Communications}, vol.~66, no.~5, pp. 2266--2280, 2018.

\bibitem{survey_1}
G.~J. {Sutton}, J.~{Zeng}, R.~P. {Liu}, W.~{Ni}, D.~N. {Nguyen}, B.~A.
  {Jayawickrama}, X.~{Huang}, M.~{Abolhasan}, Z.~{Zhang}, E.~{Dutkiewicz}, and
  T.~{Lv}, ``Enabling technologies for ultra-reliable and low latency
  communications: From phy and mac layer perspectives,'' \emph{IEEE
  Communications Surveys \& Tutorials}, vol.~21, no.~3, pp. 2488--2524, 2019.

\bibitem{fair_1}
S.~{Timotheou} and I.~{Krikidis}, ``Fairness for non-orthogonal multiple access
  in 5g systems,'' \emph{IEEE Signal Processing Letters}, vol.~22, no.~10, pp.
  1647--1651, 2015.

\bibitem{cluster_1}
M.~S. {Ali}, H.~{Tabassum}, and E.~{Hossain}, ``Dynamic user clustering and
  power allocation for uplink and downlink non-orthogonal multiple access
  ({NOMA}) systems,'' \emph{IEEE Access}, vol.~4, pp. 6325--6343, 2016.

\bibitem{fair_2}
\BIBentryALTinterwordspacing
R.~Jain, D.-M. Chiu, and W.~Hawe, ``A quantitative measure of fairness and
  discrimination for resource allocation in shared computer systems,''
  \emph{CoRR}, vol. cs.NI/9809099, 1998. [Online]. Available:
  \url{https://arxiv.org/abs/cs/9809099}
\BIBentrySTDinterwordspacing

\end{thebibliography}

\begin{IEEEbiographynophoto}{Fateme Salehi}
received the B.Sc. and M.Sc. degrees in communication engineering from University of Birjand (Birjand, Iran) in 2010 and 2012 respectively. Currently she is a Ph.D. student in communication engineering in Department of Electrical and Computer Eng. at the University of Birjand, Iran. Since March 2021, she has joined KTH Royal Institute of Technology as a visiting researcher. Her research interests include signal processing, channel estimation, MIMO-OFDM and NOMA systems, resource management in URLLC, and the Internet-of-Things.
\end{IEEEbiographynophoto}

\begin{IEEEbiographynophoto}{Naaser Neda}
received the B.S. degree in electrical Eng. from the University of Tehran and the M.S. degree in communication Eng. from Sharif University of Technology (SUT), both in Tehran, Iran, in 1990 and 1994 respectively. He received the PhD degree in communication Eng. from the University of Surrey (CCSR), Guildford, UK, in 2003. He is currently an Associate  Professor of communication engineering with the Department of Electrical and Computer Eng. at the University of Birjand, Iran. His research interests include signal processing for communication systems, physical layer of CDMA/MCCDMA/OFDM networks, sensor networks, and NOMA.
\end{IEEEbiographynophoto}

\begin{IEEEbiographynophoto}{Mohammad-Hassan Majidi}
received the B.S. degree in electrical Eng. from the Shahid Bahonar University of Kerman and the M.S. degree in communication Eng. from Imam Hossein Comprehensive University of Tehran, both in Iran, in 2003 and 2006 respectively. He received the PhD degree in telecommunication engineering from the Department of Telecommunications, Ecole Superieure d’Electricite (Supelec), Gif-sur-Yvette, France, in 2013. He is currently an Assistant Professor of communication engineering with the Department of Electrical and Computer Eng. at the University of Birjand, Iran. His research interests include signal processing for communication systems, joint channel and data detection, cryptography and secure communication.
\end{IEEEbiographynophoto}

\begin{IEEEbiographynophoto}{Hamed Ahmadi}
received the Ph.D. degree from the National University of Singapore, in 2012. He was an Agency for Science Technology and Research (A-STAR) funded Ph.D. student at the Institute for Infocomm Research (I2R), National University of Singapore. Since then, he has been working with different academic and industrial positions in Ireland and U.K. He is currently an Assistant Professor with the Department of Electronic Engineering, University of York, U.K. He is also an Adjunct Assistant Professor with the school of Electrical and Electronic Engineering, University College Dublin, Ireland. He has published more than 50 peer-reviewed book chapters, journal, and conference papers. His current research interests include design, analysis, and optimization of wireless communications networks, airborne networks, wireless network virtualization, blockchain, the Internet-of-Things, cognitive radio networks, and the application of machine learning in small cell and self-organizing networks. He is a member of Editorial Board of IEEE ACCESS, Frontiers in Blockchain, and Wireless Networks (Springer). He is a Fellow of the U.K., Higher Education Academy and Networks working Group Co-Chair and a management committee member of COST Action 15104 (IRACON).
\end{IEEEbiographynophoto}

%\newpage



\vfill


\end{document}